\documentclass[letter,longauth,traditabstract]{aa} 
\usepackage{graphicx}
\usepackage{txfonts}

\usepackage{natbib}
\bibpunct{(}{)}{;}{a}{}{,} 

\newcommand{\lsun}{\mbox{$L_\odot$}}
\newcommand{\msun}{\mbox{$M_\odot$}}

\newcommand{\micron}{$\mu$m}
\newcommand{\menv}{\mbox{$M_{\mbox{\tiny env}}$}}
\newcommand{\menvsc}{\mbox{$M_{\mbox{\tiny env}}^{< 0.05\,pc}$}}
\newcommand{\lbol}{\mbox{$L_{\mbox{\tiny bol}}$}}
\newcommand{\lsmm}{\mbox{$L_{\mbox{\tiny $\lambda> 350~\mu$m}}$}}

\begin{document}
   \title{\emph{Herschel}\thanks{\emph{Herschel} is an ESA space observatory with science
    instruments provided by European-led Principal Investigator consortia and with important
    participation from NASA.} observations of embedded protostellar clusters in the
    Rosette Molecular Cloud\thanks{Figures 5-10 are only available in electronic form via
    http://www.edpsciences.org}}
  \author{M.~Hennemann\inst{\ref{cea}},
      F.~Motte\inst{\ref{cea}},
      S.~Bontemps\inst{\ref{cea}}\fnmsep\inst{\ref{lab}},
      N.~Schneider\inst{\ref{cea}},
      T.~Csengeri\inst{\ref{cea}},
      Z.~Balog\inst{\ref{mpia}},
      J.~Di Francesco\inst{\ref{can}},
      A.~Zavagno\inst{\ref{lam}},
      Ph.~Andr\'e\inst{\ref{cea}},
      A.~Men'shchikov\inst{\ref{cea}},
      A.~Abergel\inst{\ref{ias}},
      B.~Ali\inst{\ref{nhsc}},
      J.-P.~Baluteau\inst{\ref{lam}},
      J.-Ph.~Bernard\inst{\ref{toul}},
      P.~Cox\inst{\ref{iram}},
      P.~Didelon\inst{\ref{cea}},
      A.-M.~di~Giorgio\inst{\ref{ifsi}},
      M.~Griffin\inst{\ref{cardiff}},
      P.~Hargrave\inst{\ref{cardiff}},
      T.~Hill\inst{\ref{cea}},
      B.~Horeau\inst{\ref{cea}},
      M.~Huang\inst{\ref{naoc}},
      J.~Kirk\inst{\ref{cardiff}},
      S.~Leeks\inst{\ref{ral}},
      J.~Z.~Li\inst{\ref{naoc}},
      A.~Marston\inst{\ref{hsc}},
      P.~Martin\inst{\ref{cita}},
      S.~Molinari\inst{\ref{ifsi}},
      Q.~Nguyen~Luong\inst{\ref{cea}},
      G.~Olofsson\inst{\ref{stockholm}},
      P.~Persi\inst{\ref{iasf}},
      S.~Pezzuto\inst{\ref{ifsi}},
      D.~Russeil\inst{\ref{lam}},
      P.~Saraceno\inst{\ref{ifsi}},
      M.~Sauvage\inst{\ref{cea}},
      B.~Sibthorpe\inst{\ref{ukatc}},
      L.~Spinoglio\inst{\ref{ifsi}},
      L.~Testi\inst{\ref{eso}},
      D.~Ward-Thompson\inst{\ref{cardiff}},
      G.~White\inst{\ref{ral}}\fnmsep\inst{\ref{openuni}},
      C.~Wilson\inst{\ref{mcmaster}},
      A.~Woodcraft\inst{\ref{ukatc}}}

\institute{Laboratoire AIM, CEA/IRFU -- CNRS/INSU -- Universit\'e Paris Diderot,
CEA-Saclay, 91191 Gif-sur-Yvette cedex, France\\
\email{martin.hennemann@cea.fr} \label{cea}
\and
Laboratoire d'Astrophysique de Bordeaux, CNRS/INSU -- Universit\'e de Bordeaux,
BP 89, 33271 Floirac cedex, France \label{lab}
\and
Max-Planck-Institut f\"ur Astronomie, K\"onigstuhl 17, Heidelberg, Germany \label{mpia}
\and
National Research Council of Canada, Herzberg Institute of Astrophysics,
University of Victoria, Department of Physics and Astronomy, Victoria, Canada \label{can}
\and
Laboratoire d'Astrophysique de Marseille , CNRS/INSU -- Universit\'e de Provence,
13388 Marseille cedex 13, France \label{lam}
\and
IAS, Universit\'e Paris-Sud, 91435 Orsay, France \label{ias}
\and
NHSC/IPAC, California Institute of Technology, Pasadena, CA, USA \label{nhsc}
\and
CESR \& UMR 5187 du CNRS/Universit\'e de Toulouse, BP 4346, 31028 Toulouse cedex 4, France \label{toul}
\and
IRAM, 300 rue de la Piscine, Domaine Universitaire, 38406 Saint Martin d'H\`eres, France \label{iram}
\and
INAF-IFSI, Fosso del Cavaliere 100, 00133 Roma, Italy \label{ifsi}
\and
Cardiff University School of Physics and Astronomy, UK \label{cardiff}
\and
National Astronomical Observatories, Chinese Academy of Sciences, Beijing 100012, China
\label{naoc}
\and
Space Science and Technology Department,
Rutherford Appleton Laboratory, Chilton, Didcot OX11 0NL, UK \label{ral}
\and
Herschel Science Centre, ESAC, ESA, PO Box 78, Villanueva de la Ca\~nada, 28691 Madrid, Spain \label{hsc}
\and
CITA \& Dep. of Astronomy and Astrophysics, University of Toronto, Toronto, Canada \label{cita}
\and
Department of Astronomy, Stockholm University, AlbaNova University Center, Roslagstullsbacken 21,
10691 Stockholm, Sweden \label{stockholm}
\and
INAF-IASF, Sez. di Roma, via Fosso del Cavaliere 100, 00133 Roma, Italy \label{iasf}
\and
UK Astronomy Technology Centre, Royal Observatory Edinburgh, Blackford Hill, EH9 3HJ,
UK \label{ukatc}
\and
ESO, Karl Schwarzschild Str. 2, 85748, Garching, Germany \label{eso}
\and
Department of Physics \& Astronomy, The Open University, Milton Keynes MK7 6AA, UK
\label{openuni}
\and
McMaster University, Hamilton, Canada  \label{mcmaster}
}

\titlerunning{Embedded protostellar clusters in Rosette}

\authorrunning{M.~Hennemann et al.}

\date{Received; accepted}
 
  \abstract
   {The \emph{Herschel} OB young stellar objects survey (HOBYS) has observed the Rosette
   molecular cloud, providing an unprecedented view of its star formation activity.
   These new far-infrared data reveal a population of compact young stellar
   objects whose physical properties we aim to characterise.
   We compiled a sample of protostars and their spectral energy
   distributions
   that covers the near-infrared to submillimetre wavelength range.
   These were used to constrain key properties in the protostellar evolution, bolometric luminosity,
   and envelope
   mass and to build an evolutionary diagram.
   Several clusters are distinguished including the cloud centre, the embedded clusters in the
   vicinity of luminous infrared sources, and the interaction region.
   The analysed protostellar population in Rosette ranges from 0.1 to about 15\,\msun\
   with luminosities between 1 and 150\,\lsun, which
   extends the evolutionary diagram from
   low-mass protostars into the high-mass regime.
   Some sources lack counterparts at near- to mid-infrared wavelengths,
   indicating extreme youth.
   The central cluster and the Phelps~\&~Lada~7 cluster appear less evolved than the
   remainder of the analysed
   protostellar population.
   For the central cluster, we find indications that about 25\% of the protostars classified as Class\,I
   from near- to mid-infrared data are actually candidate Class\,0 objects.
   As a showcase for protostellar evolution, we analysed four protostars of low- to
   intermediate-mass in a single dense core,
   and they represent different evolutionary stages from
   Class\,0 to Class\,I.
   Their mid- to far-infrared spectral slopes flatten towards the Class\,I stage, and the 160 to
   70\,\micron\ flux ratio is greatest for the presumed Class\,0 source.
   This shows that the \emph{Herschel} observations characterise the earliest stages of
   protostellar evolution in detail.}

   \maketitle
%

\section{Introduction}

The HOBYS \emph{Herschel} \citep{pilbratt2010} imaging survey
\citep{motte2010}
is a key programme for studying the sites of OB star formation within a distance of 3\,kpc.
The parallel mode scan map observations, carried out with the PACS \citep{poglitsch2010}
and SPIRE \citep{griffin2010,swinyard2010} instruments, provide maps of massive
cloud complexes in the wavelength range of 70 to 500\,\micron, and
thus a census of the different stages of star formation from prestellar cores to
evolved young stellar objects (YSOs).
As an excellent example of a massive cloud complex associated with -- and under the
influence of -- an
OB star cluster (NGC\,2244), the Rosette molecular cloud was observed for HOBYS
during the \emph{Herschel} Science Demonstration Phase \citep{motte2010}.
The influence of NGC\,2244 on the cloud complex is discussed in the accompanying paper
by \citet{schneider2010}, while \citet{difrancesco2010} assesses the clump population.
For consistency with \citet{schneider2010}, the distance of Rosette adopted here is 1.6\,kpc.
Many previous studies have targeted the Rosette complex.
Ongoing star formation across the molecular cloud is traced by the presence of several
luminous IRAS sources that are associated with massive clumps
seen in CO emission \citep{cox1990,williams1995,schneider1998}.
Towards these dense regions, the embedded clusters PL1 to PL7 have been
identified by \citet{phelpslada1997} from near-infrared observations.
The list of clusters in Rosette was extended by \citet{lismith2005}, \citet{romanzuniga2008},
and \citet{poulton2008} using near-infrared and \emph{Spitzer} observations.

In this paper, we make use of the unprecedented spatial resolution and sensitivity
of \emph{Herschel}
to establish a sample of compact far-infrared sources that represent protostellar objects
and to determine their fundamental properties (luminosities and envelope masses) from
their spectral energy distribution (SED).
In the early phases of the collapse of a protostellar core and the initial accretion of matter
onto a central protostar, the SED of a Class\,0 source is
dominated by thermal emission from cold dust in the envelope
\citep[$\menv > M_\star$,][]{andre2000}.
The SED of a more evolved Class\,I source is shifted to mid-infrared wavelengths,
indicating
a comparatively less massive, hotter envelope ($\menv<M_\star$).
To investigate the evolutionary stage, the protostellar envelope mass
is usually compared to the bolometric luminosity,
which serves as a proxy for stellar mass.
The \emph{Herschel} observations for the first time cover
the peak of the
protostellar SED thus constraining the evolutionary stage of early-to-evolved YSOs.
This overcomes the previous difficulties distinguishing Class\,0 and Class\,I sources using
near- to mid-infrared data.

\section{Observations}

The Rosette molecular cloud was observed by \emph{Herschel} on October 20, 2009 in the
parallel scan map mode (scanning speed of 20\arcsec/sec) simultaneously with SPIRE at
250/350/500\,\micron\ and PACS at 70/160\,\micron.
Two perpendicular scans (consisting of parallel scanlegs interspersed by turn-arounds)
were taken to
cover a SPIRE/PACS common area of $1^\circ \times 1^\circ$.
The data are reduced with scripts developed in HIPE\footnote{HIPE
is a joint development by the Herschel Science Ground Segment Consortium, consisting of ESA,
the NASA Herschel Science Center, and the HIFI, PACS, and SPIRE consortia.}
\citep[version 2.0 for SPIRE and version 3.0 for PACS]{ott2010}.
The PACS data are deglitched from cosmic ray impacts with the HIPE second-level method and then high-pass filtered
with a scanleg filter width to preserve the extended emission up to the map size scale.
The combined scans are finally projected using the HIPE MadMAP implementation
with the noise table Invntt version 1.
For details of the SPIRE data reduction, see \citet{schneider2010}.
The maps are flux-calibrated according to the correction factors of \citet{swinyard2010} and
\citet{poglitsch2010},
compared to 2MASS to correct for a $\sim$6\arcsec\ pointing offset and a
systematic offset of $\sim$4\arcsec\ between SPIRE and PACS.
The entire set of \emph{Herschel} maps is shown in \citet{motte2010}, and
Fig.~\ref{fig_pacs} (available online) shows the 70 and 160\,\micron\ maps.

\section{Results and analysis}

\begin{figure}
   \centering
   \includegraphics[width=8.8cm]{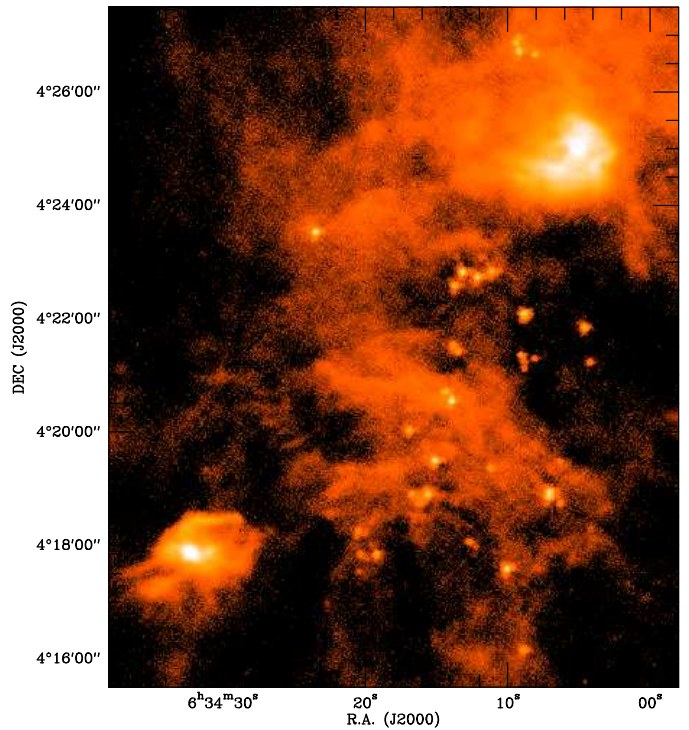}
   \caption{PACS 70\,\micron\ of the Rosette molecular cloud centre.
   The region harbours the embedded clusters PL4 (northwest), PL5 (southeast),
   and a concentration of compact \emph{Herschel} sources.}
   \label{fig_centre}
   \end{figure}

Rosette harbours several luminous far-infrared sources that represent candidate
high-mass protostars with
AFGL 961 being the brightest \citep[see][]{motte2010}.
In their vicinity, the \emph{Herschel} 70 and 160\,\micron\ maps reveal a population of
compact sources that likely
represent YSOs of low to intermediate mass.
The most prominent cluster is the one towards the Rosette molecular cloud centre
(see Fig.~\ref{fig_centre}).
The emission traces heated protostellar envelope material; in contrast, starless
cores are not expected to be detected as compact sources at 70\,\micron.
That \emph{Herschel} resolves individual protostars, even though the
Rosette region lies at an intermediate distance, allows us to study the physical
properties of the protostar population based on unprecedented far-infrared data.

\subsection{Identification of compact protostellar sources}
\label{sec_ident}

To identify the positions of the compact \emph{Herschel} protostars, 
we made use of the 70\,\micron\ \emph{MRE-GCL} catalogue \citep{motte2010} derived using
the method of \citet{motte2007}.
Spatial scales larger than 0.5\,pc were filtered out using \emph{MRE}
\citep{starckmurtagh2006}, and the \emph{Gaussclumps} programme applied \citep{kramer1998}.
We focused on the
protostellar objects clearly detected at 70\,\micron\ with $FWHM < 15\arcsec$
and derived the 70 and 160\,\micron\ flux measurements using aperture photometry.
Sources that are faint or that appear non-singular, either of which prevents
a useful measurement, were excluded.
The aperture diameters were chosen to correspond to a physical scale of 0.1\,pc (corresponding
to a source \emph{FWHM} size of roughly 0.05\,pc) and the background
is determined in adjacent annuli.
Figures~\ref{fig_centre} and \ref{fig_c4} show that protostars are resolved at
70 and 160\,\micron\ but blend together in the submillimetre, preventing a dust temperature
measurement on the protostar scale.
Thus we adopted temperatures determined by \citet{motte2010} derived from the integrated
emission on larger scales, using greybody fits with a dust emissivity
index of $\beta=2$ and assuming
$T_d = 20\,K$ for sources where no estimate is available.
This may introduce a bias towards low temperatures as the protostellar envelopes
are expected to be warmer than their surroundings.
The envelope masses were estimated by scaling the greybody curve (of dense cores)
to the 160\,\micron\ fluxes (of protostellar envelopes).
Eighty eight protostars are the basis for the further study described below.
Due to the compactness criterion, this sample does not include the high-luminosity, high-mass
protostellar objects (e.g. AFGL 961).

Partly based on \emph{Spitzer} IRAC and MIPS data of the same region obtained
by \citet{poulton2008} and on 2MASS,
a catalogue supplied by Balog~et~al. (in prep., hereafter referred to as \emph{Spitzer} catalogue)
was used to search for source
counterparts in the near- and mid-infrared.
Considering $\sim$1900 sources catalogued at 24\,\micron,
17 \emph{Herschel} sources have no counterpart.
The inspection of the maps shows that most of them are either not covered
by the MIPS observations or are not included in the catalogue owing to artifacts in the
\emph{Spitzer} maps.
Interestingly, 5 sources have only been detected with \emph{Herschel} but not
at 24\,\micron\ (completeness limit $\sim$0.5\,mJy).
The \emph{Spitzer} catalogue also includes a classification of source type
using the IRAC colours \citet{gutermuth2008}.
The compiled flux measurements that include the \emph{Spitzer} catalogue are used to estimate
bolometric luminosities
by integration over the SEDs,
where we use the submillimetre fluxes of a greybody with the adopted dust temperature.
The current lack of resolved submillimetre flux measurements means that
the mass derivation
relies on assumptions of the dust temperature and emissivity,
and we estimate that the relative accuracy
is roughly a factor of 2, and similarly for the luminosities.
This will be improved by forthcoming studies.
   \begin{figure}
   \centering
   \includegraphics[width=8.8cm]{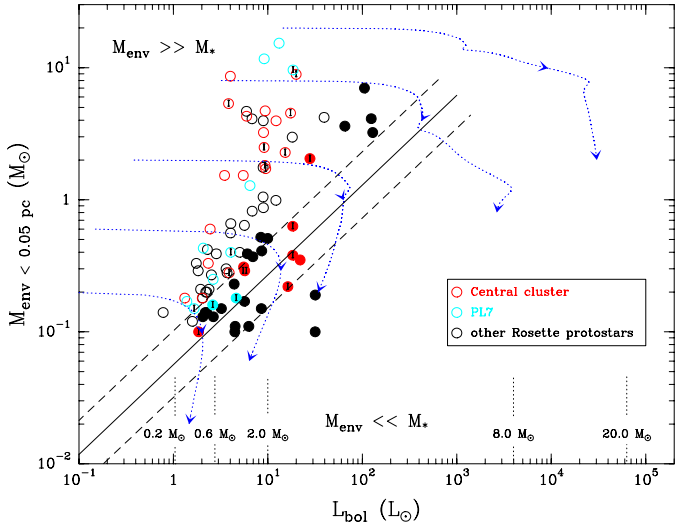}
      \caption{Envelope mass versus bolometric luminosity diagram
      for the sample of \emph{Herschel} protostars in Rosette.
      Evolutionary tracks for stellar masses
      between 0.2 and 20\,\msun\ are included.
      The solid line corresponds to 50\% of the mass accreted; the dashed lines account for
      the estimated uncertainties of \menv\ and \lbol.
      Open symbols: candidate Class\,0 protostars with \mbox{$\lsmm/\lbol>1\%$};
      filled symbols: Class\,I protostars with \mbox{$\lsmm/\lbol\leq1\%$}.}
         \label{fig_diag}
   \end{figure}

\subsection{Evolutionary stage of the Herschel protostars in Rosette}

The derived envelope masses and bolometric luminosities are plotted in an evolutionary
diagram shown in Fig.~\ref{fig_diag}.
For clarity, we omit the aforementioned error estimates.
Such a diagram is proposed to trace the evolution of embedded protostars
\citep{bontemps1996,saraceno1996}
because Class\,0 sources display relatively high envelope masses that decrease
during the mass accretion, while the luminosities (protostellar core and accretion) increase
significantly.
Evolutionary tracks are displayed for stars of different masses
\citep[cf.][]{andre2008,bontemps2010}.
The Rosette protostar sample occupies the low- to high-mass regimes in the diagram.
Compared to findings in the nearby Aquila star-forming region \citep{bontemps2010}, the
Rosette protostars are comprised of higher masses.
The diagonal lines in Fig.~\ref{fig_diag} (see caption) indicate an approximate
border zone between
envelope-dominated Class\,0 and star-dominated Class\,I objects
based on the comparison of \menv\ to $M_\star$ \citep[cf.][]{andremontmerle1994}.

A surprisingly large fraction of our sample ($\sim$2/3) falls into the candidate Class\,0 regime
above these lines.
A practical criterion inferred from this diagram is \mbox{$\lsmm/\lbol>1\%$} for Class\,0
\citep{andre2000}.
In Fig.~\ref{fig_diag} we apply this criterion obtained from the
previously compiled SEDs.
It results in more intermediate-mass objects being
classified as Class\,I, i.e., a more conservative Class\,0 assignment that we hereafter refer
to as ``candidate Class\,0".
The classification of objects lying near the border zone remains tentative because of
the uncertainty in converting the \mbox{\menv/$M_\star$} ratio into measurable quantities.
In particular, the \emph{Herschel} photometry of nearby protostars 
will reduce this uncertainty, after data from several regions have been analysed.
Until a detailed analysis of the completeness is available, our finding thus remains
preliminary.
Nevertheless, it indicates that \emph{Herschel} allows us to significantly extend the
sample of known Class\,0 objects.

\begin{table}
\caption{Protostar subsamples seen by \emph{Herschel} in Rosette}             
\label{tab_regions}      
\centering                          
\begin{tabular}{l l c}        
\hline\hline                 
Region & Associated Clusters & \# protostars \\    
\hline                        
Centre & PL4/5, REFL08, E & 27 \\      
Shell   & PL1, A                      &  5 \\      
Monoceros Ridge/ & & \\
\quad Extended Ridge & PL2, C & 11 \\      
PL3    &  -                          &  4  \\      
AFGL\,961 & PL6               &  6 \\      
PL7    & G                          & 11 \\
Distributed & -                    & 24 \\
\hline                                   
\end{tabular}
\tablefoot{Cluster names are PL for \citet{phelpslada1997},
REFL for \citet{romanzuniga2008}, A\dots G for \citet{poulton2008}.}
\end{table}

Seven protostar subsamples are established
according to their location
(field denominations in \citealt{schneider2010}),
which are listed in
Table~\ref{tab_regions}.
The distributed protostars contains sources located towards the tip of pillar
structures \citep[][]{schneider2010}.
Each subsample spans a wide mass range from 0.1 to about 10\,\msun.
In Fig.~\ref{fig_diag} the central cluster and PL7 subsamples are emphasised by coloured symbols.
In the 2 to 10\,\msun\ range, the central cluster harbours a significant number
of sources in the Class\,0 regime with 4 to 30\,\lsun.
Also the PL7 cluster contains 3 sources with relatively low luminosities.
This indicates that both the central cluster and the PL7 cluster are younger compared to the
remaining protostars.

\subsection{The classification of protostars in the central cluster}
\label{sec_c4}

Among the 27 \emph{Herschel} protostars in our central cluster sample,
the \emph{Spitzer} classification (see Sect.~\ref{sec_ident})
gives 12 Class\,I objects and one for Class\,II.
For PL7, 5 out of 11 sources are classified as Class\,I.
These classifications are added to the symbols in Fig.~\ref{fig_diag} as `I' and `II',
for comparison to
their location in the evolutionary diagram and the distinction by the $\lsmm/\lbol$ ratio.
Notably, about half of the sources seen as Class\,I by \emph{Spitzer} in these two subsamples
correspond to sources with $\lsmm/\lbol>1\%$ and thus represent candidate Class\,0
protostars.
They are intermixed with the unclassified sources in the diagram that lack detections in one
or several bands.
This suggests that the classification based on the near- to mid-infrared data alone has to be
partly revised in light of the \emph{Herschel} measurement of the protostellar envelope.

To overcome incompleteness in our \emph{Herschel} catalogue and to
provide a first census of how many sources classified using
near- to mid-infrared data are affected, we have focused on the central cluster.
We redefined the cluster area as a
rectangular field towards the Rosette molecular cloud centre and selected a total of 86 catalogued
24\,\micron\ sources in that field.
Three 24\,\micron\ sources were classified as probably extragalactic and are excluded.
We then inspected the \emph{Herschel} 70\,\micron\ map for counterparts.
The field and the resulting detection statistics are given in Table~\ref{tab_cc}.
The chosen field excludes PL4 and its vicinity because the extended emission there makes
source identification very uncertain.
Still there is a bias towards bright objects, and we estimate
an uncertainty of three sources in total and one source per subsample.
\begin{table}
\caption{Classification of protostars in the Rosette central cluster}
\label{tab_cc}
\centering
\begin{tabular}{l c c c c}
\hline\hline
 & \# total & \# Class\,II & \# Class\,I & \# unclass. \\
\hline
24\,\micron\ sources          & 83  & 39 &  26  & 18  \\
Visible in 70\,\micron        & 40 ($\pm$3)  & 10 ($\pm$1) &  19 ($\pm$1)  & 11 ($\pm$1)  \\
\hline
In \emph{Herschel} sample       & 22  &  1  &  12  &  9  \\
Candidate Class\,0 & 14  &  0  &  7   &  7  \\
\hline
\end{tabular}
\tablefoot{The cluster area is defined here by \mbox{$98.5^\circ<RA<98.6^\circ$} and
\mbox{$4.25^\circ<DEC<4.4^\circ$}.
The table head classifications base on near- to mid-infrared measurements.
Class\,0 candidates exhibit $\lsmm/\lbol>1\%$.}
\end{table}
Based on visual inspection, we find a 70\,\micron\ detection rate of about 50\%
for YSOs seen at 24\,\micron.
Due to their rising SED, we expect more 70\,\micron\ detections for Class\,I than for Class\,II,
and find about 3/4 compared to about 1/4.
In the second row of Table~\ref{tab_cc}, we list the corresponding numbers for the
analysed \emph{Herschel} sample
and the classification for the same field.
Roughly, we include about half of the visible 70\,\micron\ sources in the \emph{Herschel} list.
About 2/3 of these
are Class\,0 candidates.
This applies to 7 out of 18 unclassified sources, which indicates that many of the latter
are Class\,0 protostars.
About 25\% of 26 previously classified Class\,I sources are also Class\,0 candidates,
possibly more.
Notably, our statistical basis is low and will be improved in forthcoming studies.

   \begin{figure}
   \centering
   \includegraphics[width=8cm]{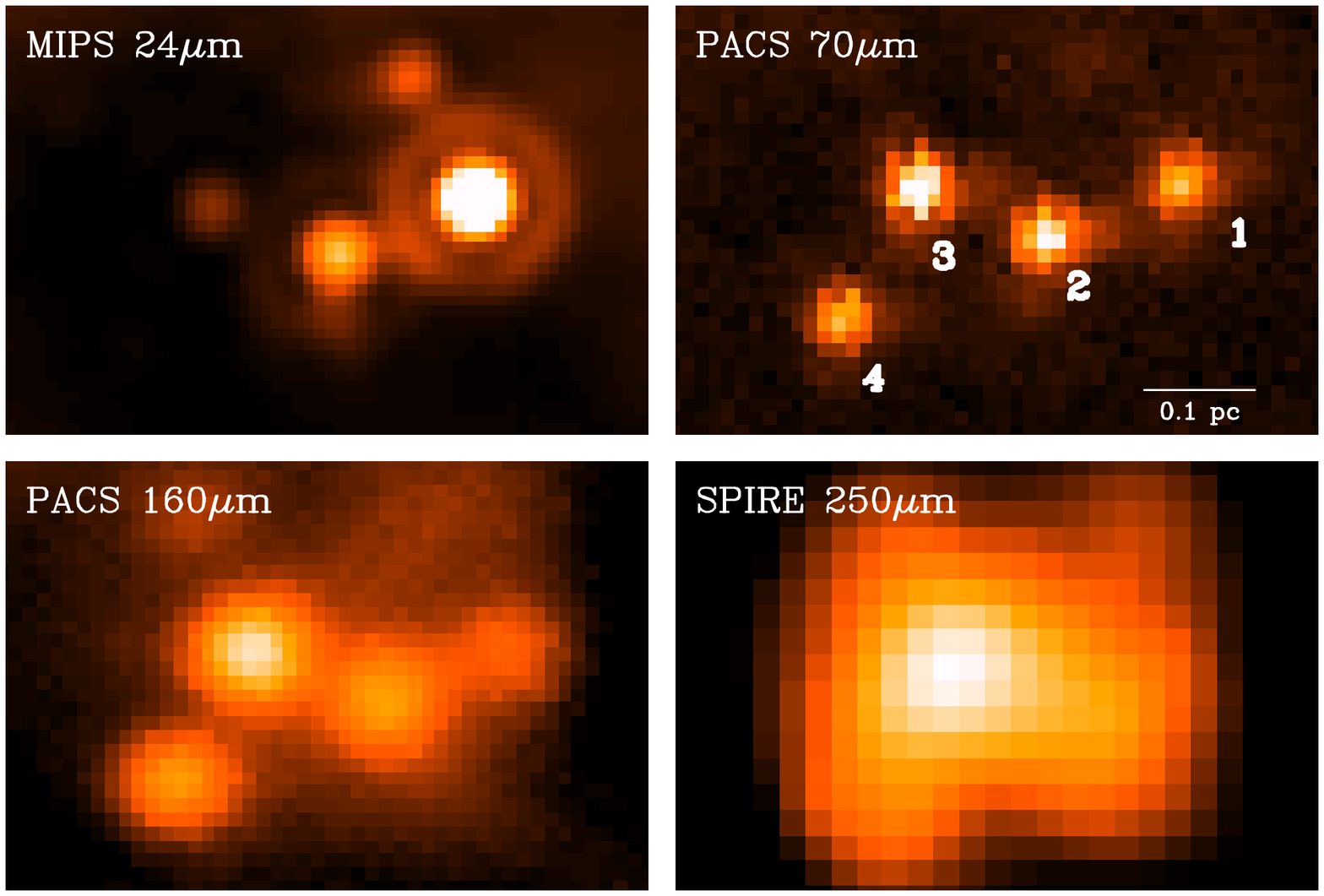}
   \includegraphics[width=8cm]{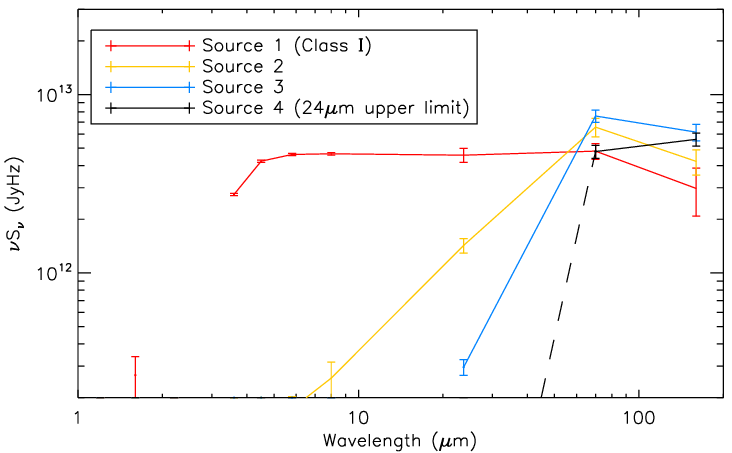}
      \caption{Four protostars in the central cluster at 24, 70, 160, and 250\,\micron\
      and their spectral energy distributions.}
         \label{fig_c4}
   \end{figure}
For illustration, we consider
a particularly interesting large dense core revealed by \emph{Herschel}
which is resolved into several protostars at 70/160\,\micron\ (Fig.~\ref{fig_c4}).
For the whole core, \citet{motte2010} derive a dust temperature of 15\,K.
We focus on 4 sources clearly detected at 70\,\micron\ labelled 1 to 4 from west to east.
The projected separation between two neighbours is around 0.14\,pc.
At 24\,\micron, only sources 1 to 3 are detected with decreasing brightness.
Based on the \emph{Spitzer} data, source 1 is classified as Class\,I, while the
others remain unclassified owing to non-detection in one or more of the IRAC bands
(sources 2 and 3)
or non-detection in all near- to mid-infrared bands (source 4).
At 70 and 160\,\micron\ source 4 shows up and the adjacent source 3 is the
brightest of the group.
The SED slopes (Fig.~\ref{fig_c4}) between 8 and 70\,\micron\ for sources 1, 2, and 3
become increasingly steep.
Beyond 70\,\micron\ they are shallower, and
the 160 to 70\,\micron\ flux ratio is the highest for source 4.
The mid- to far-infrared colours allow us to distinguish the source evolutionary stages.

The derived luminosities (\menvsc, $\lsmm/\lbol$) are
15\,\lsun\ (2\,$\msun$, 1\%) for source 1, 8\,\lsun\ (3\,$\msun$, 3\%) for source 2,
8\,\lsun\ (5\,$\msun$, 4\%) for source 3,
and 5\,\lsun\ (4\,$\msun$, 6\%) for source 4.
We note that the assumption of a single dust temperature means the relative mass differences
are not well constrained.
Relatively similar in mass, the sources represent successive evolutionary stages from
early Class\,0 (source 4) to ``flat-spectrum" Class\,I (source 1).
This is a first example of how \emph{Herschel} resolves the early evolution of protostars
in combination with \emph{Spitzer} measurements.

\section{Implications}

This initial study of the protostellar population in Rosette using the HOBYS
observations already shows that \emph{Herschel}
provides detailed insight into the earliest stages of the protostellar evolution at
low to high masses, and in particular identifies thus far elusive Class\,0 objects.
The now available \emph{Herschel} photometry will be the basis for establishing
firm criteria (mid- to far-infrared colours) for distinguishing the evolutionary
stages, including the observations
of nearby regions where individual (low-mass) protostars are resolved in the submillimetre.
Based on a complete protostar catalogue of the Rosette complex,
we will examine the evolution of the star formation activity over the whole cloud also with
respect to triggering \citep[cf.][]{schneider2010}.

\onlfig{4}{
\begin{figure*}
\includegraphics[angle=270,width=\textwidth]{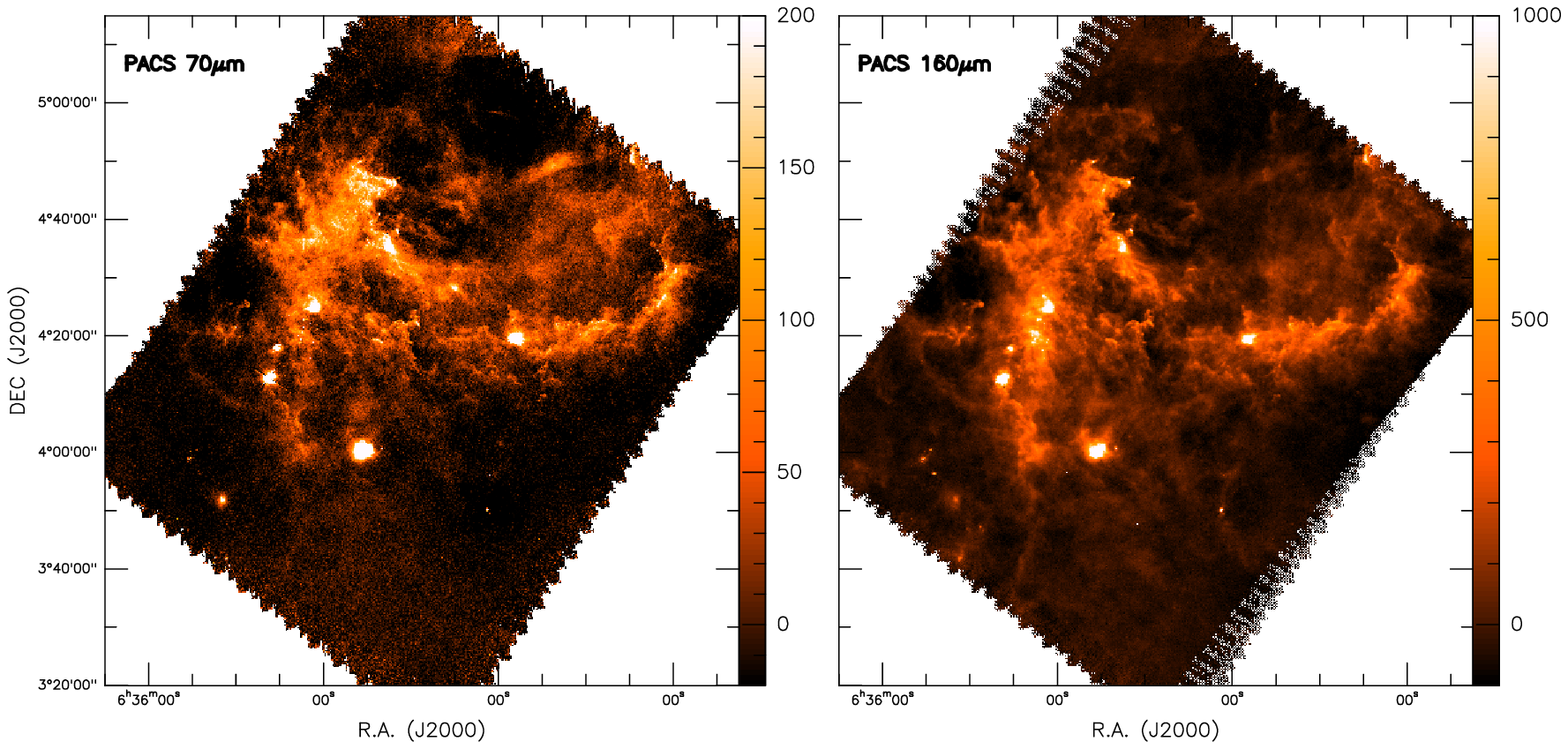}
\caption{HOBYS \emph{Herschel} 70 and 160\,\micron\ maps of the Rosette molecular cloud
(flux unit: MJy/sr). See \citet{motte2010} for details on the sensitivity of the entire
\emph{Herschel} data set.
} \label{fig_pacs}
\end{figure*}
}

\begin{acknowledgements}
SPIRE has been developed by a consortium of institutes led by
Cardiff Univ. (UK) and including Univ. Lethbridge (Canada);
NAOC (China); CEA, LAM (France); IFSI, Univ. Padua (Italy);
IAC (Spain); Stockholm Observatory (Sweden); Imperial College
London, RAL, UCL-MSSL, UKATC, Univ. Sussex (UK); Caltech, JPL,
NHSC, Univ. Colorado (USA). This development has been supported
by national funding agencies: CSA (Canada); NAOC (China); CEA,
CNES, CNRS (France); ASI (Italy); MCINN (Spain); SNSB (Sweden);
STFC (UK); and NASA (USA). 
PACS has been developed by a consortium of institutes led by MPE (Germany)
and including UVIE (Austria); KU Leuven, CSL, IMEC (Belgium); CEA, LAM
(France); MPIA (Germany); INAF-IFSI/OAA/OAP/OAT, LENS, SISSA (Italy);
IAC (Spain). This development has been supported by the funding agencies
BMVIT (Austria), ESA-PRODEX (Belgium), CEA/CNES (France), DLR (Germany),
ASI/INAF (Italy), and CICYT/MCYT (Spain).
Part of this work was supported by the ANR (\emph{Agence Nationale pour la Recherche})
project ``PROBeS", number ANR-08-BLAN-0241.
We thank the anonymous referee for helpful comments.
\end{acknowledgements}

\bibliographystyle{aa}

\end{document}